\documentstyle[12pt,epsf]{article}

  \def\pp{{\mathchoice
          {
              \kern 1pt%
              \raise 1pt
              \vbox{\hrule width5pt height0.4pt depth0pt
                    \kern -2pt
                    \hbox{\kern 2.3pt
                          \vrule width0.4pt height6pt depth0pt
                          }
                    \kern -2pt
                    \hrule width5pt height0.4pt depth0pt}%
                    \kern 1pt
           }
            {
              \kern 1pt%
              \raise 1pt
              \vbox{\hrule width4.3pt height0.4pt depth0pt
                    \kern -1.8pt
                    \hbox{\kern 1.95pt
                          \vrule width0.4pt height5.4pt depth0pt
                          }
                    \kern -1.8pt
                    \hrule width4.3pt height0.4pt depth0pt}%
                    \kern 1pt
            }
            {
              \kern 0.5pt%
              \raise 1pt
              \vbox{\hrule width4.0pt height0.3pt depth0pt
                    \kern -1.9pt  
                    \hbox{\kern 1.85pt
                          \vrule width0.3pt height5.7pt depth0pt
                          }
                    \kern -1.9pt
                    \hrule width4.0pt height0.3pt depth0pt}%
                    \kern 0.5pt
            }
            {
              \kern 0.5pt%
              \raise 1pt
              \vbox{\hrule width3.6pt height0.3pt depth0pt
                    \kern -1.5pt
                    \hbox{\kern 1.65pt
                          \vrule width0.3pt height4.5pt depth0pt
                          }
                    \kern -1.5pt
                    \hrule width3.6pt height0.3pt depth0pt}%
                    \kern 0.5pt
            }
        }}

  \def\mm{{\mathchoice
                       {
                             \kern 1pt
               \raise 1pt    \vbox{\hrule width5pt height0.4pt depth0pt
                                  \kern 2pt
                                  \hrule width5pt height0.4pt depth0pt}
                             \kern 1pt}
                       {
                            \kern 1pt
               \raise 1pt \vbox{\hrule width4.3pt height0.4pt depth0pt
                                  \kern 1.8pt
                                  \hrule width4.3pt height0.4pt depth0pt}
                             \kern 1pt}
                       {
                            \kern 0.5pt
               \raise 1pt
                            \vbox{\hrule width4.0pt height0.3pt depth0pt
                                  \kern 1.9pt
                                  \hrule width4.0pt height0.3pt depth0pt}
                            \kern 1pt}
                       {
                           \kern 0.5pt
             \raise 1pt  \vbox{\hrule width3.6pt height0.3pt depth0pt
                                  \kern 1.5pt
                                  \hrule width3.6pt height0.3pt depth0pt}
                           \kern 0.5pt}
                       }}

\catcode`@=11
\def\un#1{\relax\ifmmode\@@underline#1\else
        $\@@underline{\hbox{#1}}$\relax\fi}
\catcode`@=12

\let\du=\du                     

\def\a{\alpha}
\def\b{\beta}

\def\d{\delta}

\def\f{\phi}
\def\g{\gamma}
\def\h{\eta}

\def\k{\kappa}
\def\l{\lambda}

\def\o{\omega}
\def\p{\pi}
\def\q{\theta}
\def\r{\rho}

\def\t{\tau}

\def\x{\xi}
\def\z{\zeta}

\def\F{\Phi}

\def\L{\Lambda}
\def\O{\Omega}

\def\ve{\varepsilon}

\def\cd{{\cal D}}

\def\cf{{\cal F}}
\def\cg{{\cal G}}

\def\cj{{\cal J}}
\def\ck{{\cal K}}

\def\ct{{\cal T}}

\def\cz{{\cal Z}}

\def\bo{{\raise-.5ex\hbox{\large$\Box$}}}               
\def\pa{\partial}                                       
\def\TH{{\raise.2ex\hbox{$\displaystyle \bigodot$}\mskip-4.7mu \llap H \;}}
\def\face{{\raise.2ex\hbox{$\displaystyle \bigodot$}\mskip-2.2mu \llap {$\ddot
        \smile$}}}                                      

\def\sp#1{{}^{#1}}                              
\def\slash#1{\rlap{\hbox{$\mskip 1 mu /$}}#1}      
\def\Bar#1{\overline{#1}}                       
\def\sbar#1{\stackrel{*}{\Bar{#1}}}             

\def\VEV#1{\left\langle #1\right\rangle}        
\def\abs#1{\left| #1\right|}                    
\def\leftrightarrowfill{$\mathsurround=0pt \mathord\leftarrow \mkern-6mu
        \cleaders\hbox{$\mkern-2mu \mathord- \mkern-2mu$}\hfill
        \mkern-6mu \mathord\rightarrow$}
\def\dvec#1{\vbox{\ialign{##\crcr
        \leftrightarrowfill\crcr\noalign{\kern-1pt\nointerlineskip}
        $\hfil\displaystyle{#1}\hfil$\crcr}}}           
\def\dt#1{{\buildrel {\hbox{\LARGE .}} \over {#1}}}     

\def\frac#1#2{{\textstyle{#1\over\vphantom2\smash{\raise.20ex
        \hbox{$\scriptstyle{#2}$}}}}}                   
\def\sfrac#1#2{{\vphantom1\smash{\lower.5ex\hbox{\small$#1$}}\over
        \vphantom1\smash{\raise.4ex\hbox{\small$#2$}}}} 
\def\bfrac#1#2{{\vphantom1\smash{\lower.5ex\hbox{$#1$}}\over
        \vphantom1\smash{\raise.3ex\hbox{$#2$}}}}       
\def\afrac#1#2{{\vphantom1\smash{\lower.5ex\hbox{$#1$}}\over#2}}    

\def\[{\lfloor{\hskip 0.35pt}\!\!\!\lceil}
\def\]{\rfloor{\hskip 0.35pt}\!\!\!\rceil}

\def\du#1#2{_{#1}{}^{#2}}

\def\fracm#1#2{\hbox{\large{${\frac{{#1}}{{#2}}}$}}}

\def\tr{{\rm tr}}

\def\un{\underline}
\def\fracmm#1#2{{{#1}\over{#2}}}

\def\low#1{{\raise -3pt\hbox{${\hskip 0.75pt}\!_{#1}$}}}

\def\Dot#1{\buildrel{_{_{\hskip 0.01in}\bullet}}\over{#1}}
\def\dt#1{\Dot{#1}}

\newskip\humongous \humongous=0pt plus 1000pt minus 1000pt
\def\caja{\mathsurround=0pt}
\def\eqalign#1{\,\vcenter{\openup2\jot \caja
        \ialign{\strut \hfil$\displaystyle{##}$&$
        \displaystyle{{}##}$\hfil\crcr#1\crcr}}\,}
\newif\ifdtup

\def\ref#1{$\sp{#1)}$}

\def\pl#1#2#3{Phys.~Lett.~{\bf {#1}B} (19{#2}) #3}
\def\np#1#2#3{Nucl.~Phys.~{\bf B{#1}} (19{#2}) #3}

\def\cqg#1#2#3{Class.~and Quantum Grav.~{\bf {#1}} (19{#2}) #3}
\def\cmp#1#2#3{Commun.~Math.~Phys.~{\bf {#1}} (19{#2}) #3}

\def\ibid#1#2#3{{\it ibid.}~{\bf {#1}} (19{#2}) #3}

\parskip=\medskipamount                 
\thicklines                         

\begin{document}


\thispagestyle{empty}               

\def\border{                                            
        \setlength{\unitlength}{1mm}
        \newcount\xco
        \newcount\yco
        \xco=-24
        \yco=12

        \par\vskip-8mm}

\def\headpic{                                           
        \indent
        \setlength{\unitlength}{.8mm}
        \thinlines
        \par
        \par\vskip-6.5mm
        \thicklines}

\border\headpic {\hbox to\hsize{
\vbox{\noindent  NBI--HE--00--17 \hfill March 2000 \\ 
ITP--UH--04/00 \hfill   hep-th/0003245 \\
DESY 00 -- 048 \hfill  }}}

\noindent
\vskip1.3cm
\begin{center}

{\Large\bf  Anomalous N=2 Superconformal Ward Identities}

\vglue.3in

Sergei V. Ketov~\footnote{Also at HCEI, Academy of Sciences, Akademichesky 4,
Tomsk 634055, Russia} 

{\it Niels Bohr Institute}\\
{\it University of Copenhagen}\\
{\it 2100 Copenhagen $\slash{O}$, Denmark}\\

and

{\it Institut f\"ur Theoretische Physik, Universit\"at Hannover}\\
{\it Appelstra\ss{}e 2, D--30167, Hannover, Germany}\\
{\sl ketov@itp.uni-hannover.de}

\end{center}

\vglue.3in

\begin{center}
{\Large\bf Abstract}
\end{center}

The N=2 superconformal Ward identities and their anomalies are discussed in 
N=2 superspace (including N=2 harmonic superspace), at the level of the 
low-energy effective action (LEEA) in four-dimensional N=2 supersymmetric 
field theories. The (first) chiral N=2 supergravity compensator is related to 
the known N=2 anomalous Ward identity in the N=2 (abelian) vector mulitplet 
sector. As regards the hypermultiplet LEEA given by the N=2 non-linear 
sigma-model (NLSM), a new anomalous N=2 superconformal Ward identity is found,
 whose existence is related to the (second) analytic compensator in N=2 
supergravity. The celebrated solution of Seiberg and Witten is known to obey 
the (first) anomalous Ward identity in the Coulomb branch. We find a few 
solutions to the new anomalous Ward identity, after making certain assumptions
about unbroken internal symmetries. Amongst the N=2 NLSM target space metrics 
governing the hypermultiplet LEEA are the $SU(2)$-Yang-Mills-Higgs monopole 
moduli-space metrics that can be encoded in terms of the spectral curves 
(Riemann surfaces), similarly to the Seiberg-Witten-type solutions. After a
dimensional reduction to three spacetime dimensions (3d), our results support 
the mirror symmetry between the Coulomb and Higgs branches in 3d, N=4 gauge 
theories.

\noindent

\newpage

\section{Introduction}

Field theories with N=2 extended supersymmetry are known to possess remarkable 
properties that sometimes allow one to obtain their exact (non-perturbative) 
low-energy solutions, pioneered by Seiberg and Witten \cite{sw}. The natural 
way to produce such exact results is to relate the field theory problem to an 
integrable system (or Whitham dynamics in the Seiberg-Witten case) \cite{mmm}.
The origin of an elliptic curve behind the exact solution \cite{sw} also 
becomes apparent in this approach. Another (related) interpretation of the 
Seiberg-Witten result is possible from the viewpoint of anomalous breaking of 
N=2 superconformal symmetry and uniformization theory on Riemann surfaces 
\cite{mat}. It is not very surprising since N=2 superconformal symmetry is 
well-known to be instrumental in deriving the classical structure of all 
(non-conformal) N=2 supersymmetric field theories, including N=2 supergravity,
in four spacetime dimensions. It is natural to examine first the constraints 
imposed by N=2 superconformal invariance and then study compensation of 
unwanted N=2 superconformal symmetries. Or one can start from a classical 
field theory that is N=2 superconformally invariant, and then investigate its 
superconformal anomalies that can be developed in quantum field theory. 

Much work in the recent past was devoted to investigating the constraints 
imposed by N=4 superconformal symmetry on the correlation functions of the N=4 
super-Yang-Mills theory in the context of AdS/CFT correspondence  \cite{mal}. 
More recently, similar constraints on the correlation functions were studied 
in the context of four-dimensional N=2 conformal supersymmetry \cite{sok}. We
recall that all N=2 supersymmetric field theories can be brought into the 
manifestly N=2 supersymmetric form, by using {\it off-shell} (unconstrained) 
N=2 superfields in {\it Harmonic Superspace} (HSS) \cite{gikos}. It makes the 
effective field theory methods to be very efficient in the N=2 case contrary 
to the N=4 case where only on-shell N=4 supersymmetry is possible. The next 
obvious step is to make manifest N=2 superconformal invariance of the N=2 
supersymmetric quantum effective action, and then study its superconformal 
anomalies. Truly non-perturbative results are expected to be obtained along 
these lines. 

As regards quantum field theories with rigid N=2 supersymmetry, their building
 blocks are given by N=2 vector multiplets and hypermultiplets. At the level 
of the {\it Low-Energy Effective Action} (LEEA), on the (abelian) N=2 vector 
multiplet side we have to deal with the Seiberg-Witten-type action specified by
a holomorphic potential and the associated special K\"ahler geometry. This 
topic is well-known, and we are going to use it as our basic pattern to follow.
 Hypermultiplets (e.g., if they are magnetically charged) may also develop the
non-trivial LEEA that takes the form of a hyper-K\"ahler 
{\it Non-Linear Sigma-Model} (NLSM) by N=2 supersymmetry. After being 
formulated in HSS, the N=2 NLSM also possess an (analytic) potential. Our 
purpose in this paper is to impose and make manifest N=2 superconformal 
invariance in the N=2 NLSM, and then formulate an anomalous N=2 superconformal 
Ward identity on the hyper-K\"ahler potential.

The paper is organized as follows. In sect.~2 we briefly review the N=2
supercurrents, the anomalous N=2 superconformal Ward identities in the N=2 
gauge sector, and their relation to the solution of Seiberg and Witten 
\cite{sw}. In sect.~3 we construct in HSS the most general N=2 superconformal 
NLSM that gives a general solution to the special hyper-K\"ahler geometry. The 
anomalous N=2 superconformal Ward identity on the hypermultiplet LEEA 
(N=2 NLSM) is found for the first time in sect.~4, by using the N=2 
supergravity compensators in HSS. In sect.~5 we give our conclusions. 
In Appendix we briefly review a construction of N=2 supergravity in HSS.

\section{N=2 supercurrent and Ward identities}

An N=2 supercurrent is the irreducible representation of N=2 supersymmetry in
four spacetime dimensions, having superspin one. The 
independent field components of the N=2 supercurrent (of some N=2 matter 
system) include the energy-momentum tensor, the N=2 supersymmetry current, the
central charge current, the axial current, the $SU(2)$ current of R-symmetry 
and some auxiliary field components of lower dimension. 

The relevant field components of the N=2 supercurrent were first identified in
 ref.~\cite{sohn} by analyzing the free field theory of a massive 
(Fayet-Sohnius) hypermultiplet. The systematic way of derivation of the N=2 
supercurrent superfield is provided by a construction of the irreducible N=2 
superfields in the conventional (flat) N=2 superspace 
$\{\cz\}=(x^m,\q_{\a}^i,\bar{\q}^{\dt{\a}}_j)$, where $m=0,1,2,3$, $\a=1,2$ 
and 
$i,j=1,2$. All irreducible N=2 superprojectors were found in ref.~\cite{sg}, 
whereas all irreducible (scalar) N=2 superfields were explicitly derived in 
ref.~\cite{kty}. Amongst the irreducible N=2 superfields, comprising a general
N=2 real scalar superfield, one finds almost all off-shell N=2 supermultiplets 
that usually appear in any discussion of N=2 supersymmetry (with a finite 
number of the auxiliary fields). In particular, an N=2 (restricted) chiral 
superfield $\F(x,\q,\bar{\q})$ is defined by the N=2 superspace (off-shell) 
constraints
$$ \bar{D}^i_{\dt{\a}}\F=0~,\quad D^4 \F=\bo\bar{\F}~,\eqno(2.1)$$ 
where $(D^{\a}_i,\bar{D}_{\dt{\a}}^j)$ are the usual (flat) N=2 superspace 
covariant derivatives.~\footnote{We use the notation 
$D_{\a\b}=D_{i\a}D^i_{\b}$, $D^{ij}=D^{i\a}D^j_{\a}$ and 
$D^4=\fracmm{1}{12}D_{ij}D^{ij}$, and similarly for the \newline ${~~~~~}$ 
conjugated quantities.} As a consequence of the constraints (2.1), the N=2
restricted chiral superfield $\F$ possess a two-form $F=F_{mn}dx^m\wedge dx^n$ 
satisfying the `Bianchi identity' $dF=0$. A solution to the `Bianchi identity',
 $F=dA$,  in terms of the one-form $A$ subject to the gauge transformations 
$\d A=d\l$, allows one to represent the N=2 superfield strength $W$ of an 
abelian N=2 vector multiplet by a restricted chiral N=2 superfield too. 

The N=2 restricted chiral superfield $\F$ is dual to an
 N=2 linear superfield $L^{ij}$. The latter is symmetric with respect to its 
$SU(2)$ indices and satisfies the off-shell constraints
$$ D^{(i}\low{\a}L^{jk)}=\bar{D}{}^{(i}_{\dt{\a}}L^{jk)}=0~,\quad
\Bar{(L^{ij})}=\ve_{ik}\ve_{jl}L^{kl}~.\eqno(2.2)$$
The duality relation in N=2 superspace is just given by
$$ L^{ij}=D^{ij}\F~.\eqno(2.3)$$
Both superfields $\F$ and $L^{ij}$ represent the irreducible N=2 multiplets of
superspin zero and superisospin zero. 

Similarly, an irreducible N=2 scalar superfield $R$ of superspin zero and 
superisospin one is defined by the constraints \cite{kty} 
$$ D\low{\a\b}R=\bar{D}_{\dt{\a}\dt{\b}}R=i\[D\low{\a}^j,\bar{D}_{j\dt{\a}} \]R
=0~.\eqno(2.4)$$
The irreducible N=2 scalar superfield $R$  is dual to an N=2 projective 
superfield $T^{ijkl}$,
$$ T^{ijkl}=D^{(ij}\bar{D}^{kl)}R~,\eqno(2.5)$$
where $T^{ijkl}$ is totally symmetric with respect to its $SU(2)$ indices 
and obeys the (off-shell) constraints
$$ D^{(i}\low{\a}T^{jklm)}=\bar{D}{}^{(i}_{\dt{\a}}T^{jklm)}=0~,\quad
\Bar{(T^{i_1i_2i_3i_4})}=\ve_{i_1j_1}\ve_{i_2j_2}\ve_{i_3j_3}\ve_{i_4j_4}
T^{j_1j_2j_3j_4}~.\eqno(2.6)$$
The N=2 supercurrent $J$ is also in the list of the irreducible N=2 scalar
superfields, being defined by the constraints \cite{kty}
$$ D_{ij}J=\bar{D}_{ij}J=0~.\eqno(2.7)$$
The N=2 superspace constraints (2.7) imply that the energy-monentum tensor
is symmetric, conserved and traceless, whereas all the vector currents are
conserved. In other words, $J$ is a multiplet of N=2 superconformal currents.

An N=2 superconformal anomaly amounts to breaking the N=2 supercurrent 
conservation relations (2.7) by an N=2 anomaly multiplet of lower superspin, 
i.e. of superspin zero. This is equivalent to activating an irreducible 
superspin-zero superfield in the N=2 scalar superfield $J$. As is clear 
from the above discussion, there are potentially {\it two} ways of assigning 
the N=2 anomaly multiplet with an N=2 irreducible superspin-zero superfield:
either $\F$ (or, equivalently, $L^{ij}$) or $R$ (or, equivalently, $L^{ijkl}$).
The main difference between the two choices is the fact that $L^{ij}$ still 
contains a conserved vector current (associated with unbroken central charge 
transformations), whereas  the vector current in $L^{ijkl}$ is not conserved.
The first choice yields the N=2 superconformal anomaly relation in the 
standard form \cite{westb} 
$$  \fracm{i}{4}D_{ij}J= L_{ij}~.\eqno(2.8)$$
For example, the N=2 supercurrent conservation law in the quantum N=2 SYM 
theory takes the form \cite{westb,marcu}
$$ \fracm{i}{4}D_{ij}J=\bar{D}_{ij}\bar{S}~,\qquad S=\fracmm{c}{2}\tr W^2~,
\eqno(2.9)$$
where the (Lie algebra-valued) N=2 SYM superfield strength $W$ has been 
introduced. The constant $c$ is proportional to the one-loop renormalization 
group beta-function. Though $\tr W^2$ is merely a chiral (not a restricted 
chiral) N=2 superfield, eq.~(2.9) can be easily brought into the form (2.8) 
by a local shift of the supercurrent, $J\to J-4iS$.    

The most general N=2 supersymmetric {\it Ansatz} for the LEEA of some number 
$(r)$ of abelian N=2 vector multiplets is governed by a holomorphic potential
$\cf(W)$ of the N=2 (restricted chiral) superfield strengths $W_p$ 
\cite{poten, poten1}~,\footnote{Our normalization differs from that used in 
ref.~\cite{sw} by a factor $-i/(16\p)$.}
$$ I[W]=\int d^4x d^4\q\, \cf(W_p) + {\rm h.c.}\eqno(2.10)$$
The superfield $W$ has conformal weight $+1$, in accordance with its canonical
dimension, whereas the N=2 chiral superspace measure in eq.~(2.10) has 
conformal weight $(-2)$. It is, therefore, clear that the N=2 superconformal
Ward identity for the LEEA (2.10) is given by
$$ \sum^r_{p=1} W_p\fracmm{\pa \cf}{\pa W_p} -2\cf=0~.\eqno(2.11)$$ 
The N=2 superconformal solution to the holomorphic potential $\cf(W)$ is thus 
given by a homogeneous (of degree two) function. The (rigid) N=2 
superconformal invariance of the action (2.10) is the necessary pre-requisite 
for its coupling to N=2 conformal supergravity \cite{poten}.  

Given a non-trivial renormalization flow (like in the N=2 SYM theory), the 
N=2 superconformal Ward identity (2.11) is going to be broken by the 
anomaly $S$,
$$\sum^r_{p=1} W_p\fracmm{\pa \cf}{\pa W_p} -2\cf=4S~.\eqno(2.12)$$  
This equation is just the anomalous N=2 superconformal Ward identity for 
(abelian) N=2 vector multiplets, which was found in ref.~\cite{hw} by 
`averaging' the anomaly relation (2.9) with respect to the quantum effective 
action (2.10). A simple derivation of eq.~(2.12) by using the N=2 supergravity 
compensators is discussed in sect.~4. 

Equation (2.12) can be applied to a derivation of the Seiberg-Witten solution
\cite{sw} provided that one knows the anomaly $S$ as the function of $W$, in 
the context of the N=2 SYM theory based on the gauge group $SU(2)$ 
spontaneously broken to $U(1)$. The original derivation \cite{sw} made use of 
the electric-magnetic duality and renormalization flow. Since the anomaly $S$
is N=2 chiral, gauge-invariant and of dimension two, its vacuum expectation 
value has to be proportional to the order parameter 
$u=\fracm{1}{2}\VEV{\tr W^2}$, with the coefficient being dictated by the 
one-loop beta-function $\b_1$ --- see eq.~(2.9). A comparision with 
ref.~\cite{sw} yields \cite{hw} 
$$ c=2\p i\b_1~. \eqno(2.13)$$
To close eq.~(2.12), as the equation on $\cf$, one needs a relation 
between $a=\VEV{W}$ and $u$. It was obtained in ref.~\cite{mat} by using the
modular invariance of $u=u(a)$, in the form of a non-linear differential 
equation,
$$ (1-u^2)u'' + \frac{1}{4}au'{}^3=0~.\eqno(2.14)$$
A connection to integrable systems arises after identifying the
moduli space of the Coulomb branch in the Seiberg-Witten model with the 
moduli space of complex structures on an elliptic curve. The Seiberg-Witten 
solution then appears to be a classical solution to the equations of motion 
of a particular spin chain system \cite{mmm}, while the origin of the elliptic
curve underlying the dynamics becomes apparent in this approach.~\footnote{The
origin of the elliptic curve in the Seiberg-Witten exact solution is also 
explained by brane \newline ${~~~~~}$ technology in the context of M-theory 
\cite{fourw}.} Generalizations to the larger gauge groups and the presence 
of N=2 matter are straightforward, in principle.

Our main goal, however, is to explore what can happen on the hypermultiplet 
side, at the level of the LEEA. Unlike the N=2 vector multiplets, the universal
and most symmetric off-shell formulation of hypermultiplets is only possible 
in HSS, with the infinite number of the auxiliary fields.
 
\section{Rigid N=2 superconformal symmetry in HSS} 

In the HSS approach \cite{gikos} the standard N=2 superspace coordinates
$\{\cz\}=(x^m,\q_{\a}^i,\bar{\q}^{\dt{\a}}_j)$ are extended by bosonic 
harmonics 
(or twistors) $u^{\pm i}$, $i=1,2$, belonging to the group $SU(2)$ and 
satisfying the unimodularity condition
$$ u^{+i}u^-_i=1~,\quad \Bar{u^{i+}}=u^-_i~.\eqno(3.1)$$
The hidden analyticity structure of the N=2 superspace constraints defining 
both N=2 vector multiplets and FS hypermultiplets, as well as their solutions 
in terms of unconstrained N=2 superfields, can be made manifest in HSS 
\cite{giosbook}.

Instead of an explicit parametrization of the twistor sphere 
$S^2=SU(2)/U(1)$, the $SU(2)$-covariant HSS approach deals with the 
equivariant functions of harmonics, having the definite $U(1)$ charges defined
by $U(u^{\pm}_i)=\pm 1$. The simple harmonic integration rules,
$$ \int du =1 \quad{\rm and}\quad \int du\, u^{+(i_1}\cdots
u^{+i_m}u^{-j_1} \cdots
u^{-j_n)}=0 ~~{\rm otherwise}~,\eqno(3.2)$$
are similar to the (Berezin) integration rules in superspace. In particular, 
any harmonic integral over a $U(1)$-charged quantity vanishes. The harmonic 
covariant derivatives, preserving the defining equations (3.1) in the original
(central) basis, are given by
$$ \pa^{++}=u^{+i}\fracmm{\pa}{\pa u^{-i}}~,\quad 
\pa^{--}=u^{-i}\fracmm{\pa}{\pa u^{+i}}~,\quad
\pa^{0}=u^{+i}\fracmm{\pa}{\pa u^{+i}}-u^{-i}\fracmm{\pa}{\pa u^{-i}}~~.
\eqno(3.3)$$
They satisfy an $su(2)$ algebra and commute with the standard (flat) N=2
superspace covariant derivatives $D^{\a}_i$ and $\bar{D}_{\dt{\a}}^j\,$. The 
operator $\pa^0$ measures $U(1)$ charges.

The key feature of HSS is the existence of the {\it analytic} subspace 
parametrized by $$ (\z^M;u)=\left\{ \begin{array}{c}
x^{\a\dt{\a}}_{\rm analytic}=x^{\a\dt{\a}}-4i\q^{i\a}\bar{\q}^{\dt{\a}j}
u^+_{(i}u^-_{j)}~,~~
\q^+_{\a}=\q^i_{\a}u^+_i~,~~ \bar{\q}^+_{\dt{\a}}=\bar{\q}^i_{\dt{\a}}u^+_i~;~~
u^{\pm}_i \end{array} \right\}~,\eqno(3.4)$$
which is invariant under N=2 (rigid) supersymmetry \cite{gikos}:
$$ \d x^{\a\dt{\a}}_{\rm analytic}=-4i\left( \ve^{i\a}\bar{\q}^{\dt{\a}+}
+\q^{\a+}\bar{\ve}^{\dt{\a}i}\right)u^-_i
\equiv -4i\left( \ve^{\a-}\bar{\q}^{\dt{\a}+}
+\q^{\a+}\bar{\ve}^{\dt{\a}-}\right)~,$$
$$ \d\q^+_{\a}=\ve^i_{\a}u^+_i\equiv \ve^+_{\a}~,\quad
\d\bar{\q}^+_{\dt{\a}}=\bar{\ve}^i_{\dt{\a}}u^+_i\equiv 
\bar{\ve}_{\dt{\a}}{}^+~,\quad \d u_i^{\pm}=0~.\eqno(3.5)$$
The analytic dependence includes $\q^+_{\hat{\a}}$ but not $\q^-_{\hat{\a}}$, 
where $\hat{\a}=(\a,\dt{\a})$.

The usual complex conjugation does not preserve analyticity. However, it does,
after being combined with another (star) conjugation that only acts on the
$U(1)$ indices as $(u^+_i)^*=u^-_i$ and $(u^-_i)^*=-u^+_i$. One has 
$ \sbar{u^{\pm i}}=-u^{\pm}_i$ and $\sbar{u^{\pm}_i}=u^{\pm i}$.

{\it Analytic} superfields $\f^{(q)}(\z(\cz,u),u)$ of any positive (integral) 
$U(1)$ charge $q$ in HSS are defined by the off-shell constraints ({\it cf.} 
 the definition of N=1 chiral superfields)  
$$D^+\low{\a}\f^{(q)}=\bar{D}^+_{\dt{\a}}\f^{(q)}=0~,\quad {\rm where}\quad
D^{+}_{\a}=D^i_{\a}u^+_i \quad {\rm and}\quad
\bar{D}^+_{\dt{\a}}=\bar{D}^i_{\dt{\a}}u^+_i~.\eqno(3.6)$$
The analytic measure reads $d\z^{(-4)}du\equiv d^4x^{m}_{\rm analytic}
d^2\q^+d^2\bar{\q}^+du$, and it is of $U(1)$ charge $(-4)$. The covariant 
derivatives in the analytic basis (3.4) receive certain connection terms. For
example, the harmonic derivative $\pa^{++}$ in the analytic subspace is 
replaced by  
$$D^{++}=\pa^{++}-4i\q^{\a+}\bar{\q}^{\dt{\a}+}
\fracmm{\pa}{\pa x^{\a\dt{\a}}}~.\eqno(3.7)$$
This derivative preserves analyticity and permits integration by parts. 
Similarly, one easily finds the $U(1)$ charge operator in the analytic 
subspace reads
$$D^{0}=u^{+i}\fracmm{\pa}{\pa u^{+i}}
-u^{-i}\fracmm{\pa}{\pa u^{-i}}+\q^{\a+}\fracmm{\pa}{\pa\q^{\a+}}+
\bar{\q}^{\dt{\a}+}\fracmm{\pa}{\pa\bar{\q}^{\dt{\a}+}}~~.\eqno(3.8)$$
In what follows we always use the analytic basis and the associated HSS 
covariant derivatives denoted by capital $D$, without making explicit 
references.

The use of harmonics also gives us control over the (linearly realised) 
$SU(2)_R$ symmetry (or its absence), in the context of manifest N=2 
supersymmetry (see ref.~\cite{rec} for more details). Since the translational
and Lorentz symmetries, as well as N=2 supersymmetry, are manifestly realized 
in HSS, the latter provides us with the natural arena for a study of 
`truly' N=2 superconformal symmetries on the top of N=2 non-conformal 
 (rigid or Poincar\'e) supersymmetry.

The superfield transformation rules with respect to dilatations (with the 
infinitesimal parameter $\r$) are dictated by conformal weights $w$ of the 
superfields, together with the weights of the N=2 superspace coordinates,
$$ w[x]=1~,\quad w[\q]=w[\bar{\q}]=\fracm{1}{2}~,\quad w[u]=0~.\eqno(3.9)$$
The non-trivial part of the N=2 superconformal transformations is given by 
$SU(2)_{\rm conf.}$ internal rotations with the parameters $l^{ij}$, special 
conformal transformations with the parameters $k_{\a\dt{\a}}\,$, and N=2 
special supersymmetry with the parameters $\h^i_{\a}$ and 
$\bar{\h}^{\dt{\a}}_i\,$.

The N=2 superconformal extension of the spacetime conformal transformations,
$$ \d x^{\a\dt{\a}}=\r x^{\a\dt{\a}}+k_{\b\dt{\b}}x^{\a\dt{\b}}x^{\b\dt{\a}}~,
\eqno(3.10)$$
is dictated by the requirement of preserving the unimodularity and analyticity
 conditions in eqs.~(3.1) and (3.6), respectively. As regards the non-trivial
part of the N=2 superconformal transformation laws, one finds \cite{giosc}
$$ \eqalign{
\d x^{\a\dt{\a}} & ~=~ -4i\l^{ij}u^-_iu^-_j\q^{\a+}\bar{\q}^{\dt{\a}+}
+k_{\b\dt{\b}}x^{\a\dt{\b}}x^{\b\dt{\a}} 
+4i\left( x^{\a\dt{\b}}\bar{\q}^{\dt{\a}+} \bar{\h}^-_{\dt{\b}}
-x^{\dt{\a}\b}\q^{\a+}\h^-_{\b} \right)~,\cr
\d\q^{\a+}& ~=~ \l^{ij}u^+_iu^-_j\q^{\a+} + k_{\b\dt{\b}}x^{\a\dt{\b}}\q^{\b+}
-2i (\q^{\b+}\q^+_{\b}) \h^{\a-} + x^{\a\dt{\b}}\bar{\h}^+_{\dt{\b}}~,\cr
\d\bar{\q}^{\dt{\a}+} & ~=~ -\sbar{(\d\q^{\a+})}~,\cr
\d u^+_i& ~=~ \left[ \l^{kj}u^+_ku^+_j 
+4ik_{\a\dt{\a}}\q^{\a+}\bar{\q}^{\dt{\a}+}
+4i\left( \q^{\a+}\h^+_{\a}+\bar{\h}^+_{\dt{\a}}\bar{\q}^{\dt{\a}+}\right)
\right] u^-_i~,\cr
\d u^-_i& ~=~ 0~.\cr}\eqno(3.11)$$
Since the building blocks of invariant actions in HSS are given by the 
measure, analytic superfields and HSS covariant derivatives, only their
transformation properties under the rigid `truly' N=2 superconformal 
transformations are needed. It follows from eq.~(3.11) that \cite{giosc}
$${\rm Ber} \fracmm{\pa(\z',u')}{\pa(\z,u)}= 1-2\L~,\quad{\rm or}
\quad \d[d\z^{(-4)}du]=-2\L[d\z^{(-4)}du]~,\eqno(3.12)$$
where the HSS superfield parameter
$$\L=-\left(\r + k_{\a\dt{\a}}x^{\a\dt{\a}}\right) +
\left(\l^{ij}+4i\q^{\a i}\h^j_{\a}+4i\bar{\h}^j_{\dt{\a}}\bar{\q}^{\dt{\a}i}
\right)u^+_iu^-_j \eqno(3.13)$$
has been introduced. Similarly, one easily finds that
$$ (D^{++})'=D^{++} - (D^{++}\L)D^0\quad {\rm and}\quad (D^0)'=D^0~.
\eqno(3.14)$$

The truly (rigid) N=2 superconformal infinitesimal parameters can, therefore, 
be encoded into the single scalar harmonic superfield $\L$ that is subject to 
the constraint \cite{giosc}
$$ (D^{++})^2\L=0~,\eqno(3.15)$$
and the reality condition
$$ \sbar{(\L^{++})}=\L^{++}~,\quad {\rm where}\quad \L^{++}\equiv D^{++}\L~.
\eqno(3.16)$$
The transformations rules of the harmonics,
$$ \d u^+_i=\L^{++}u^-_i~,\quad \d u^-_i=0~,\eqno(3.17)$$
together with eqs.~(3.12), (3.14), (3.15) and (3.16) yield the very simple and 
 convenient description of rigid N=2 conformal supersymmetry (on the top of 
N=2 Poincar\'e supersymmetry) in N=2 HSS.

The {\it special} hyper-K\"ahler geometry of the N=2 (rigidly) superconformal 
NLSM in components was investigated in ref.~\cite{wkv}. A general solution to
the special hyper-K\"ahler geometry in HSS was described in our recent paper 
\cite{mine}. We use the pseudo-real $Sp(1)$ notation for a {\it Fayet-Sohnius}
(FS) hypermultiplet superfield,
$$ q^+_a=(\sbar{q}{}^+,~q^+)~,\quad a=1,2~, \quad q^{a+}=\ve^{ab}q^+_b~,
\eqno(3.18)$$
which can be easily generalized to the case of several FS hypermultiplets, 
$q^{a+}\to q^{A+}$ and $q^+_A=\O_{AB}q^{B+}$, with a constant (antisymmetric) 
$Sp(k)$-invariant metric $\O_{AB}$,  $A,B=1,\ldots,2k$.

First, we recall that the most general (rigidly) N=2 supersymmetric NLSM can 
be formulated in terms of the FS hypermultiplet superfields,
$$ I_{\rm NLSM}[q] =-\fracmm{1}{\k^2}
\int d\z^{(-4)}du \left[ \fracm{1}{2}q_A^{+}D^{++}q^{A+} 
+\ck^{(+4)}(q^{A+},u^{\pm}_i)\right]~,\eqno(3.19)$$ 
where the real analytic function $\ck^{(+4)}=\sbar{\ck^{(+4)}}$ of $U(1)$ 
charge $(+4)$ is known as a hyper-K\"ahler (pre-)potential 
\cite{giosk}.~\footnote{The HSS superfields $q$ are dimensionless. The 
dimensionality of the measure in the action (3.19) \newline ${~~~~~}$ is 
compensated by the coupling constant $\k$ of dimension of length.}  
By manifest N=2 supersymmetry of the NLSM action (3.19), the NLSM metric must 
be hyper-K\"ahler for any choice of $\ck^{(+4)}$. Unfortunately, an explicit
general relation between a hyper-K\"ahler potential and the corresponding 
hyper-K\"ahler metric is not available (see, however, 
refs.~\cite{giosk,met,mtn} for the explicit hyper-K\"ahler potentials of (ALF)
multi-Taub-NUT and Atiyah-Hitchin metrics, and their derivation from the NLSM 
(3.19) in HSS, and ref.~\cite{mybook} for a review or a general introduction
into the supersymmetric NLSM).    

Eq.~(3.19) formally solves the hyper-K\"ahler constraints on the NLSM metric
in terms of an arbitrary function $\ck^{(+4)}$ that may be considered as the
(analytic) hypermultilet counterpart to the (holomorphic) potential $\cf$ of 
abelian N=2 vector superfields in eq.~(2.10). It is, therefore, natural to 
impose extra N=2 superconformal invariance on the action (3.19), in order to 
determine a general solution to the special hyper-K\"ahler geometry, since the
 free part of the action (3.19) is N=2 superconformally invariant \cite{giosc}.
 The FS superfields $q^+$ have conformal weight one, $\d q^+=\L q^+$. Together
 with eqs.~(3.12), (3.17) and (3.19) it implies two constraints \cite{mine}:
$$ \fracmm{\pa\ck^{(+4)}}{\pa q^{A+}}q^{A+}=2\ck^{(+4)}\quad {\rm and}\quad
 \fracmm{\pa\ck^{(+4)}}{\pa u^+_i}=0~.\eqno(3.20)$$
This means that the special hyper-K\"ahler potentials are given by 
{\it homogeneous (of degree two)} functions $\ck^{(+4)}(q^{A+},u^-_i)$ of 
$q^{A+}$. There is no restriction on the dependence of $\ck^{(+4)}$ upon 
$u^-_i$, though there should be no dependence upon $u^+_i$. 

The HSS description of the N=2 superconformal {\it hypermultiplet} actions 
depending upon the FS superfields in terms of a homogeneous (degree 2) 
potential is thus formally the same as that of the N=2 superconformal 
(abelian) vector multiplet actions in the standard N=2 (chiral) superspace 
(sect.~2). However, the special K\"ahler geometry in the target space of the 
NLSM arising in the scalar sector of the N=2 superconformal action (2.10) is
very different from the special hyper-K\"ahler geometry arising from the N=2 
superconformal NLSM action (3.19) in components.

\section{Ward identities and supergravity compensators}

The  best  way of derivation of the superconformal anomaly relations is 
based on the use of the {\it supergravity} (SG) compensators \cite{books}. 
To compensate the unwanted (local) N=2 superconformal symmetries in N=2 
conformal SG, one needs two compensators \cite{van}. This also implies the 
existence of {\it two} anomaly relations in the N=2 case ({\it cf.} sect.~2).

In the case of N=2 superfield SG, its most universal formulation is provided 
by HSS \cite{giosc,giosu,kuz} --- see Appendix for a short introduction and 
our notation. Having an N=2 matter action $I$ coupled to N=2 SG, 
one can naturally {\it define} an N=2 supercurrent $\cj$ by a 
variation of the action $I$ with respect to the N=2 conformal SG potential 
$\cg$,
$$ \cj=\fracmm{\d I}{\d\cg}~,\eqno(4.1)$$
in the flat limit where all N=2 conformal SG fields vanish. The N=2 conformal 
SG potential $\cg$ is defined by eq.~(A.4), where it is introduced as the 
general N=2 {\it harmonic} real superfield, subject to the pre-gauge 
transformations (A.5) and the gauge transformations (A.7) whose linearized 
form reads \cite{kuz}
$$ \d\cg= D^{++} l^{--}~.\eqno(4.2)$$

The N=2 superconformal anomalies are also naturally defined in HSS by 
variational derivatives of the N=2 matter action with respect to the N=2 SG 
 compensators $v^{++}_5(\z,u)$ and $\o(\z,u)$ \cite{kuz},
$$ L^{++}=\fracmm{\d I}{\d v^{++}_5}~,\eqno(4.3)$$
and
$$\ct^{++++}=\fracmm{\d I}{\d\o}~,\eqno(4.4)$$
where $v^{++}_5$ is the real {\it analytic} gauge superfield, associated with 
the central charge and transforming under the (linearized) gauge 
transformations (A.13) as 
$$ \d v^{++}_5= D^{++}\l_5~,\eqno(4.5)$$
whereas the {\it analytic} density $\o$ transforms according to eq.~(A.14) that
implies (in the linearized approximation)
$$ \d\o= -D^{--}\L^{++}= -D^{--}(D^+)^4 l^{--}~,\eqno(4.6)$$
where we have used the last eq.~(A.6). The HSS superfields, $L^{++}$ and 
$\ct^{(+4)}\equiv\ct^{++++}$, representing the N=2 superconformal anomalies,
are analytic by their definition. 

Given the N=2 matter action $I$ that is entirely formulated in the ordinary
N=2 superspace  without harmonics (it is certainly the case in the N=2 gauge 
sector, without FS hypermultiplets), one can make a connection to the standard
N=2 anomaly relation (2.8). The invariance of the action $I$ with 
respect to the gauge transformations (4.2) obviously yields 
$$ D^{++}\cj=0~,\eqno(4.7)$$
whereas the invariance of the same action with respect to the gauge
transformations (4.5) implies 
$$ D^{++}L^{++}=0~.\eqno(4.8)$$
Equation (4.7) means that $\cj$ is independent upon harmonics too, whereas 
eq.~(4.8) implies that $L^{++}=u^+_iu^+_jL^{ij}(x,\q,\bar{\q})$, where 
$L^{ij}$ satisfies the N=2 linear multiplet constraints (2.2) due to the 
analyticity of $L^{++}$. Finally, the invariance of the action $I$ with 
respect to the pre-gauge transformations (A.5) and (A.11) yields \cite{kuz}
$$ \fracm{i}{4}(D^+)^2\cj= L^{++}~, \eqno(4.9)$$
which is equivalent to eq.~(2.8), as expected. 

The HSS results about N=2 SG potentials and compensators imply the natural 
definitions of the latter in the ordinary N=2 superspace by linear relations,
$$ G=\int du \,\cg(\z,u,\bar{\q}^-) \eqno(4.10)$$
and
$$ \F=\int du (\bar{D}^-)^2v^{++}_5(\z,u)~.\eqno(4.11)$$
The N=2 real superfield $G(x,\q,\bar{\q})$ gives the N=2 conformal 
SG potential ({\it cf.} ref.~\cite{hst}), whereas the abelian N=2 
(restricted chiral) superfield strength squared, $\F^2$, can serve as an N=2 
(unrestricted) chiral density, i.e. as the N=2 chiral compensator 
 ({\it cf.} ref.~\cite{hw}). The anomaly relation (2.8) is then the 
direct consequence of the definitions
$$ J=\fracmm{\d I}{\d G}\quad {\rm and}\quad
 S=\fracmm{\d I}{\d \F^2}~.\eqno(4.12)$$ 

Having identified the compensator $\F$ with one of the (abelian) N=2 gauge 
superfield strengths, $\F=W_1$, taking the N=2 gauge LEEA (2.10) to represent
the N=2 matter action $I$ above results in the anomalous N=2 superconformal 
Ward identity (2.12). The second compensator decouples from the effective 
N=2 gauge matter action, so that it does not have any impact on its anomaly 
structure. The situation is just the opposite one in the case where the 
effective N=2 matter action represents a selfinteraction of FS 
hypermultiplets. The analytic compensator $\o$ can be considered as a part 
(density) of the hypermultiplet matter, while the invariance of the most 
general N=2 matter action in HSS with respect to the gauge transformations 
(4.2) and (4.6) gives rise to the second anomaly relation in the form  
\cite{kuz}
$$ D^{++}\cj = D^{--}\ct^{(+4)}~.\eqno(4.13)$$
Being applied to the hypermultiplet LEEA $I$ in the form of the N=2 NLSM 
(3.19) in HSS, eq.~(4.13) gives rise to the following anomalous N=2 
superconformal Ward identity:
$$ \sum_A q^{A+}\fracmm{\pa\ck^{(+4)}}{\pa q^{A+}}-2\ck^{(+4)}=\ct^{(+4)}~.
\eqno(4.14)$$
This is the key equation in our paper. A generic anomaly $\ct^{(+4)}(q,u)$ is
analytic, while it has to be invariant under the unbroken gauge symmetries,
internal symmetries, and modular transformations, if any. The anomalous Ward 
identity (4.14) then gives us the equation on the hyper-K\"ahler potential 
$\ck^{(+4)}$ of the effective N=2 NLSM (LEEA).

\section{Examples}

To `close' the anomalous N=2 superconformal Ward identity (4.14), as the 
equation on the effective hyper-K\"ahler potential $\ck(q,u)$, one has to know 
the anomaly $\ct$ as a function of $(q,u)$ explicitly. In the absence of a 
general solution for the anomaly (at least, we are unaware ot it), it is 
worthy to discuss some examples. The non-anomalous symmetries play the major 
r\^ole in determining the form of the anomaly $\ct(q,u)$.

The crucial simplification arises when the $SU(2)_R$ automorphisms of N=2 
supersymmetry algebra are not broken (together with the N=2 supersymmetry that
we always assume). Since the $SU(2)_R$ transformations are linearly realised 
in HSS, the R-invariance of the hypermultilet LEEA amounts to the independence 
of the anomaly $\ct$ (and, hence, of the hyper-K\"ahler potential $\ck$) upon 
harmonics. Since both have $U(1)$ charge $(+4)$, the most general (analytic) 
invariant `Ansatz' is given by a {\it real quartic polynomial} of the analytic
 FS superfields $q^{A+}$ \cite{rec},
$$\ct^{(+4)}(q)\sim \ck^{(+4)}(q)=\l_{ABCD}q^{A+}q^{B+}q^{C+}q^{D+}~,
\eqno(5.1)$$
whose coefficients $\l_{(ABCD)}$ are totally symmetric, being subject to the 
reality condition,  $\sbar{\ck}{}^{(+4)}=\ck^{(+4)}$. Not all of the 
coefficients in eq.~(5.1) are really significant since the FS kinetic terms in 
eq.~(3.19) have the manifest global $Sp(n)$ symmetry. It may be not accidental
that this $Sp(n)$ symmetry coincides with the maximal $Sp(n)$ holonomy group 
of the hyper-K\"ahler manifolds in $4n$ real dimensions.

In the case of a single FS hypermultiplet, eq.~(5.1) is simplified to
$$ \ct^{(+4)}\sim \ck^{(+4)}= \fracm{\l}{2}(\sbar{q}{}^+)^2(q^+)^2+\left[ 
\g \sbar{(q^+)}{}^4 +  \b\sbar{(q^+)}{}^3 q^+ +{\rm h.c.}\right] \eqno(5.2)$$
with one real $(\l)$ and two complex $(\b,\g)$ parameters. The $Sp(1)$ 
transformations of $q^+_a$ leave the form of eq.~(5.2) invariant, but not its
coefficients, which can be used to reduce the number of coupling constants in 
the family of the hyper-K\"ahler metrics described by the hyper-K\"ahler 
potential (5.2) from five to two. In addition, eq.~(5.2) implies the (on-shell)
conservation laws 
$$ D^{++}\ck^{(+4)}=D^{++}\ct^{(+4)}=0~,\eqno(5.3)$$
which are valid on the equations of motion of the hypermultiplet (FS) 
superfield,
$$D^{++}\sbar{q}{}^+ =\pa\ck^{(+4)}/\pa q^+\quad {\rm and}\quad 
D^{++}q^+ =-\pa\ck^{(+4)}/\pa\sbar{q}{}^+~.\eqno(5.4)$$

To understand the physical significance of eq.~(5.2), it is instructive 
to consider first a simpler case, by assuming the additional (translational) 
$U(1)_T$ symmetry that acts on the complex superfields $(q^+,\sbar{q}{}^+)$ by
phase rotations (with a constant parameter $\a$),
$$ q^{+}\to e^{i\a}\q^+~,\qquad \sbar{q}{}^+\to e^{-i\a}\sbar{q}{}^+~,
\eqno(5.5)$$
but does not move the hyper-K\"ahler structure in the target space of the N=2
 NLSM (3.19). It happens, e.g., in the N=2 supersymmetric QED with a single 
charged hypermultiplet, or in the Coulomb branch of the Seiberg-Witten model 
\cite{rev}. In geometrical terms, the $U(1)_T$ symmetry amounts to the 
existence of a tri-holomorphic (translational) isometry in the N=2 NLSM target 
space. It is worth mentioning that the $SU(2)_R$ isometries are not 
tri-holomorphic but rotational: they rotate three independent complex 
structures in the N=2 NLSM hyper-K\"ahler target space. Given the
$SU(2)_R\times U(1)_T$ isometry of the N=2 NLSM target space, it must be the
symmetry of the NLSM hyper-K\"ahler potential, which implies further 
restrictions in eq.~(5.2). The unique, $SU(2)_R\times U(1)_T$ invariant,
hyper-K\"ahler potential is obviously given by 
$$ \ck^{(+4)}_{\rm Taub-NUT}=\fracmm{\l}{2}\left(\sbar{q}{}^+q^+\right)^2~,
\eqno(5.6)$$
while it is known as the hyper-K\"ahler potential of the Taub-NUT metric with
the mass parameter $M=\frac{1}{2}\l^{-1/2}$ \cite{met}.

\begin{figure}
\vglue.1in
\makebox{
\epsfxsize=4in
\epsfbox{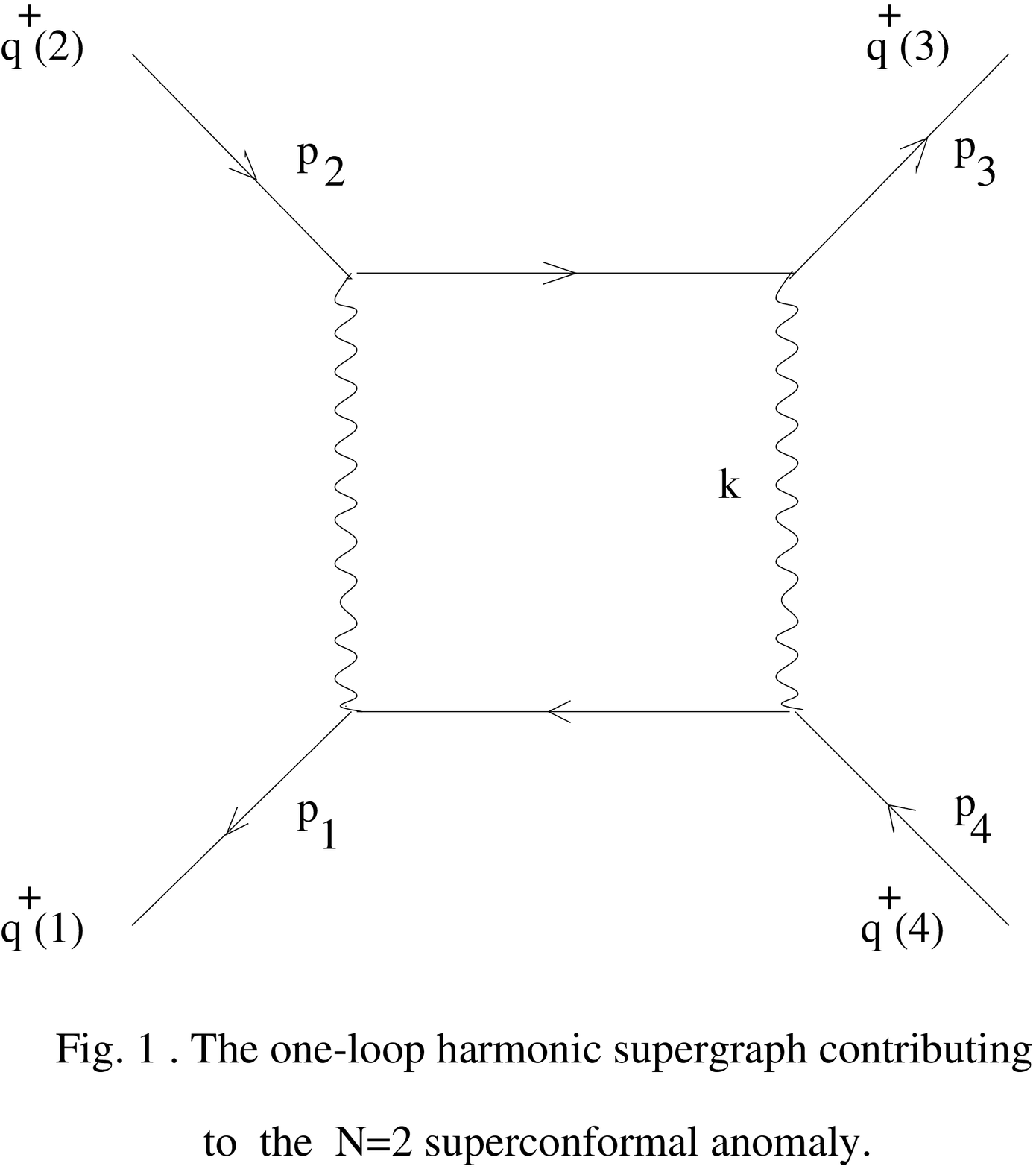}
}
\end{figure}

The induced coupling constant $\l$ in the one-loop approximation is determined
by the HSS graph depictured in Fig.~1. As was shown in ref.~\cite{ikz}, this 
graph does lead to the non-vanishing anomalous contribution  having the form 
(5.6), provided that the matter hypermultiplet has {\it a non-vanishing} 
central charge. The wave lines in Fig.~1 denote the analytic propagators of 
the N=2 (abelian) vector superfields $V^{++}$ (in N=2 supersymmetric Feynman 
gauge) \cite{hssf}, 
$$ i\VEV{ V^{++}(1)V^{++}(2)}=\fracmm{1}{\Box_1}(D_1^+)^4\d^{12}(\cz_1-\cz_2)
\d^{(-2,2)}(u_1,u_2)~,\eqno(5.7)$$
where $\d^{(-2,2)}(u_1,u_2)$ stands for the harmonic delta-function 
\cite{hssf}. The hypermultiplet analytic propagators (the solid lines in 
Fig.~1) with {\it non-vanishing} central charges are given by \cite{ikz}
$$ i\VEV{q^+(1)q^+(2)}=\fracmm{-1}{\Box_1+m^2}
\fracmm{(D^+_1)^4(D^+_2)^4}{(u^+_1u_2^+)^3}e^{\t_3[v(2)-v(1)]}\d^{12}
(\cz_1-\cz_2)~,\eqno(5.8)$$
where the `bridge' $v$ satisfies an equation ${\cal D}_Z^{++}e^v=0$, and
$m^2=\abs{Z}^2$ stands for the hypermultiplet (BPS) mass. We find \cite{ikz}
$$ \l=\fracmm{g^4}{\p^2}\left[ \fracmm{1}{m^2}\ln\left( 1+\fracmm{m^2}{\L^2}
\right) -\fracmm{1}{\L^2+m^2}\right]~, \eqno(5.9)$$
where the gauge coupling constant $g$ and the IR-cutoff $\L$ have been 
introduced. It is not difficult to check that $\l\neq 0$ only if $Z\neq 0$. 
The naive `non-renormalization theorem' usually forbids the quantum 
corrections given by the integrals over a subspace of the full N=2 superspace,
like the one in eq.~(3.19). However, this `theorem' does not apply here, 
because of the non-vanishing central charges that give rise to the  explicit 
dependence of the superpropagators (5.8) upon the N=2 superspace Grassmann 
(anticommuting) coordinates via the bridges $v(\q,\bar{\q})$ that are 
responsible for the N=2 superconformal anomaly.

The more general $SU(2)_R$-invariant anomaly (5.2) cannot be generated in N=2
perturbation theory, but it can be generated non-perturbatively, due to 
instanton contributions \cite{ah}. Of course, in an abelian N=2 supersymmetric
 field theory no instantons exist. This means that the N=2 
perturbative anomaly described by the Taub-NUT metric is exact in the abelian 
case. If, however, the underlying N=2 supersymmetric quantum field theory 
has a non-abelian gauge group of rank larger than one (say, $SU(3)$), then one
 may expect the nonperturbative contributions to the hypermultiplet LEEA 
(in the Higgs branch) from instantons and anti-instantons that break the 
$U(1)_T$ symmetry. The on-shell relations (5.3) imply that in this case both
the anomaly $\ct^{(+4)}$ and the hyper-K\"ahler potential $\ck^{(+4)}$ can be
expressed in terms of a real analytic superfield $T^{++++}$ satisfying the
constraints  
$$ D^{++}T^{++++}=0\quad {\rm and}\quad \sbar{T}{}^{++++}=T^{++++}~, 
\eqno(5.10)$$
which are {\it the same} as the off-shell constraints (2.6) defining an $O(4)$
projective superfield $T^{ijkl}$, in the ordinary N=2 superspace,
$T^{++++}(\z,u)=u^+_iu^+_ju^+_ku^+_lT^{ijkl}(x,\q,\bar{\q})$. Unlike the $O(2)$
 tensor multiplet defined by eq.~(2.2), the $O(4)$ multiplet does not have a 
conserved vector amongst its field components. Hence, the $U(1)$ isometry, if 
any, in the N=2 NLSM to be constructed in terms of $T^{++++}$, is no longer
tri-holomorphic (or translational). The Taub-NUT NLSM (5.6) arises in the 
limit $T^{++++}\to (L^{++})^2$.  

The two-parametric family of the hyper-K\"ahler potentials (5.2) describes
the (hyper-K\"ahler and $SU(2)_R$-invariant) deformations of the 
{\it Atiyah-Hitchin} (AH) metric \cite{ah}. The N=2 (projective) superspace 
description of the AH metric in terms of an $O(4)$ projective supermultiplet 
(2.6) was found in ref.~\cite{ir}. The AH metric is known to be the only 
{\it regular} metric in the family \cite{ati}. The `difference' between the 
AH and Taub-NUT metrics, being considered as the metrics in the quantum moduli
space of an N=2 gauge theory (in the region where quantum perturbation theory 
applies), can be interpreted as the (exponentially small) instanton 
contributions \cite{sw3}. Similar remarks are valid in the more 
general case (5.1) \cite{rec}.

Another simple example of the N=2 superconformal anomaly, which is still under 
control, is possible in the case of the unbroken $U(1)_T$ symmetry (5.5). It 
implies that the function $\ct^{(+4)}(q^+,\sbar{q}{}^+;u)$ be a function of 
the invariant product $(q^+\sbar{q}{}^+)$ and harmonics $u_i^-$ only. The most 
general `Ansatz' reads
$$ \ct^{(+4)}(q^+\sbar{q}{}^+;u)=\sum^{\infty}_{l=0}\x^{(-2l)}
\fracmm{(\sbar{q}{}^+q^+)^{l+2}}{l+2}~,\eqno(5.11)$$
whose harmonic-dependent `coefficients' $\x^{(-2l)}$ are given by
$$ \x^{(-2l)}=\x^{(i_1\cdots i_{2l})}u^-_{i_1}\cdots u^-_{i_{2l}}~,\qquad
l=1,2,\ldots~,\eqno(5.12)$$
and obey the reality condition
$$ \sbar{\x}{}^{(-2l)}=(-1)^l\x^{(-2l)}~.\eqno(5.13)$$
The associated solution to eq.~(4.14) takes the similar form (5.11), while it
appears to be the hyper-K\"ahler potential describing the multi-Taub-NUT 
metrics \cite{mtn}. The multi-Taub-NUT metrics are known to describe static
configurations of several (BPS) monopoles  \cite{ati}. This is not surprising
from the viewpoint of the brane engineering of the effective N=2 
supersymmetric gauge field theories in M-theory \cite{fourw}, where the N=2 
field theory hypermultiplets are associated with (parallel) D6-branes whose 
configurations in M-theory are just described by the multi-Taub-NUT metrics
\cite{tsen}.

The brane technology/M-theory also suggest a possibility of yet another 
generalization, by adding a (parallel) orientifold $O6^-$ to the D6-branes 
\cite{tsen}. In geometrical terms, this means replacing the $U(1)_T$ 
translational isometry by a rotational $U(1)_R$ isometry in the N=2 NLSM 
target space. Though the associated N=2 NLSM in HSS were recently constructed 
in terms of an $O(4)$ projective superfield \cite{mine}, we are unaware about
 their explicit reformulation in terms of the FS superfields. We conjecture
that this should give rise to the $D_k$ series of the asymptotically locally 
flat (self-dual) metrics in the four-dimensional target space of N=2 NLSM.

\section{Conclusion}

In the preceeding section we gave a few non-trivial examples of the N=2 
superconformal anomaly $\ct^{(+4)}$ that saturates the hyper-K\"ahler 
potential of the effective hypermultilet LEEA given by an N=2 NLSM. To get 
more (non-perturbative) solutions to our main eq.~(4.14), it may be better to
find first the organizing principle behind all those solutions. In the 
Seiberg-Witten case (2.12), it was the underlying Riemann surface or an 
integrable system. It is natural to expect a similar hidden curve behind the 
hypermultiplet LEEA too \cite{rec}. The $SU(2)_R$-invariant effective NLSM 
metrics in our examples (sect.~5) coincide with the standard metrics in the
(BPS) monopole moduli space of the classical $SU(2)$-Yang-Mills-Higgs system, 
with magnetic charge $n=1$ (Taub-NUT) or $n=2$ (Atiyah-Hitchin).  In the N=2 
NLSM, whose target space metric is given by the monopole moduli space metric 
of higher magnetic charge $n>2$, the $SU(2)_R$ symmetry is necessarily broken
 \cite{rec}. As was shown in ref.~\cite{hit}, a BPS monopole of magnetic 
charge $n>1$ can always be described by the {\it spectral curve} 
(Riemann surface) of genus $(n-1)^2$. It is therefore, conceivable that more 
general solutions to the anomalous N=2 superconformal Ward identity (4.14) 
are also encoded in terms of the spectral curve, like the Seiberg-Witten-type 
solutions to eq.~(2.12). After a dimensional reduction to three spacetime 
dimensions, our results support the conjectured mirror symmetry between 
the Coulomb and Higgs branches \cite{sint}.

Quantum breaking of conformal symmetry in a (classically) superconformal field 
theory is related to the appearance of dynamically generated scales. The 
superconformal compensators can be naturally interpreted as the superfield 
extensions of the scale parameters. From this physical point of view, the 
(first) N=2 chiral compensator is apparently related to the N=2 superfield 
extension of the renormalization scale squared, whereas the second N=2 
compensator appears to be the N=2 superfield extension  of the induced N=2 
NLSM coupling constant that is (roughly) proportional to the inverse central 
charge squared (sect.~5). It would be interesting to understand better the 
physical significance of the second N=2 compensator from the viewpoint of the 
underlying (non-abelian) N=2 gauge theory and brane technology.

\section*{Acknowledgements} I am indebted to L. Alvarez-Gaum\'e, J. Ambjorn, 
S. Cherkis, S. Gates Jr., A. Gorsky and E. Ivanov for useful discussions. I am
also grateful to Niels Bohr Institute in Copenhagen for hospitality extended 
to me during the completion of this work.

\newpage

\section*{Appendix: N=2 supergravity in HSS}

All (rigid) N=2 supersymmetric field theories can be naturally defined in HSS,
in terms of {\it unconstrained} analytic superfields, with manifest N=2 
supersymmetry \cite{gikos,giosbook}. N=2 matter (FS) hypermultiplets are 
described by complex analytic superfields $q^{+}$ of $U(1)$ charge $+1$, 
whereas their coupling to N=2 super-Yang-Mills fields is described via the 
(Lie algebra-valued) extension of the FS hypermultiplet kinetic operator 
$D^{++}$ by a gauge connection $V^{++}$, so that the new gauge-covariant 
operator $\cd^{++}=D^{++}+V^{++}$ preserves analyticity. The rigid N=2 
superconformal transformations also preserve the analytic subspace of flat N=2
HSS (sect.~3), so it is natural to define N=2 conformal {\it supergravity} 
(SG) in HSS along the similar lines \cite{giosc,giosu}: by preserving 
analyticity (of N=2 matter \& gauge fields), the analytic conjugation (= the 
product of usual complex conjugation and Weyl reflection of the sphere $S^2$),
and unimodularity (of harmonics). In this Appendix we briefly review the HSS 
formulation of N=2 SG along the lines of ref.~\cite{kuz} --- see 
refs.~\cite{giosbook,giosc,giosu,kuz} for more details. 

Let $\{\z^M,u,\q^{\hat{\a}-}\}$ be the coordinates of the full N=2 HSS in 
the analytic basis, $\hat{\a}=(\a,\dt{\a})$. The {\it conformal} N=2 SG 
transformations are naturally defined in HSS as the analyticity-preserving 
diffeomorphisms \cite{giosc,giosu},
$$\eqalign{
\d\z^M~=~&\l^M(\z,u)~,\cr
\d u^{i+}~=~&\l^{++}(\z,u)u^{i-}~,\quad \d u^{i-}=0~,\cr
\d\q^{\hat{\a}-}~=~&\l^{\hat{\a}-}(\z,u,\q^-)~.\cr}\eqno(A.1)$$ 
Accordingly, the covariant derivative $\cd^{++}$ in N=2 conformal SG can be put
into the form
$$ \cd^{++}=D^{++}+H^{M++}D_M+H^{(+4)}D^{--}+H^{\hat{\a}+}D^+_{\hat{\a}}~,
\eqno(A.2)$$
where the SG {\it vielbeine} $(H^{M++},H^{(+4)}\equiv H^{++++},H^{\hat{\a}+})$
have been introduced in front of the flat N=2 HSS covariant derivatives
$D_M=(\pa_{\a\dt{\a}},~D^-_{\hat{\a}})$, $D^{--}$ and $D^+_{\hat{\a}}$. The N=2
conformal SG parameters (A.1) can be similarly organized into the one-forms
$$ \l=\L^MD_M+\L^{++}D^{--} \quad {\rm and}\quad 
\r=\r^{\hat{\a}-}D^+_{\hat{\a}}~.\eqno(A.3)$$ 

Since the SG derivative $\cd^{++}$ is supposed to preserve analyticity,
$D^+_{\hat{\a}}\cd^{++}\F^{(+p)}=0$, of the analytic superfield $\F^{(+p)}$ 
of a positive $U(1)$ charge $p$, $D^+_{\hat{\a}}\F^{(+p)}=0$, the vielbeine of 
eq.~(A.2) have to obey certain linear constraints, whose solution is given by 
\cite{kuz}
$$H^{\a\dt{\a}++}=-iD^{\a+}\bar{D}^{\dt{\a}+}\cg~,\quad 
H^{\a+++}=-\fracm{1}{8}D^{\a+}(\bar{D}^+)^2\cg~,\quad
H^{(+4)}=(D^+)^4\cg~,\eqno(A.4)$$
where $(D^+)^4=\fracm{1}{16}(D^+)^2(\bar{D}^+)^2$, and the real unconstrained 
(general HSS superfield) N=2 SG pre-potential $\cg(\z,u,\q^-)$ is subject to 
the pre-gauge transformations \cite{kuz}
$$ \d \cg =\fracm{1}{4}(D^+)^2\O^{--}+ \fracm{1}{4}
(\bar{D}^+)^2\sbar{\O}{}^{--} \eqno(A.5)$$
with the complex unconstrained HSS parameter $\O^{--}(\z,u,\q^-)$. Accordingly,
 the gauge HSS superfield parameters $(\L^M,\L^{++})$ are to be  expressed 
in terms of a single real unconstrained HSS superfield $l^{--}(\z,u,\q^-)$ 
\cite{kuz},
$$\L^{\a\dt{\a}}=-iD^{\a+}\bar{D}^{\dt{\a}+}l^{--}~,\quad
\L^{\a+}=-\frac{1}{8}D^{\a+}(\bar{D}^+)^2l^{--}~,\quad 
\L^{++}=(D^+)^4l^{--}~.\eqno(A.6)$$
It is easy to see that $H^{\hat{\a}+}$ is pure gauge, so that we can simply 
ignore it, $H^{\hat{\a}+}=0$. Then the transformation law of $\cg$ under the 
remaining gauge transformations with the parameters $(\l,l^{--})$ is given by 
a simple formula, 
$$ \d \cg =-\l \cg + \cd^{++} l^{--}~.\eqno(A.7)$$  
In the absence of superconformal anomalies, there exist a (WZ-type) gauge, 
where the HSS superfield $\cg$ is independent upon harmonics, being subject 
to the constraints 
$$ D_{ij}\cg(x,\q,\bar{\q})=\bar{D}_{ij}\cg(x,\q,\bar{\q})=0~. \eqno(A.8)$$
This equation coincides with eq.~(2.7) defining the N=2 conformal 
supercurrent, so that the independent field components of $\cg$ (in the WZ 
gauge) are in one-to-one correspondence with the field content of the 
off-shell N=2 conformal SG, as it should.

To construct N=2 matter and gauge couplings in N=2 Poincar\'e (non-conformal or
Einstein) SG, one has to {\it compensate} `truly' N=2 superconformal gauge
transformations, namely, dilatations, special conformal transformations, 
$U(1)$ chiral and $SU(2)_{\rm conf.}$ rotations, and N=2 special 
supersymmetry \cite{van}. Some of them are compensated by a real analytic
gauge superfield $V^{++}_5(\z,u)$ as the {\it first compensator}, subject to 
abelian gauge transformations in HSS,
$$ \d V^{++}_5= \cd^{++}\L_5~, \eqno(A.9)$$
with the real analytic parameter $\L_5(\z,u)$. In the context of gauging the
N=2 supersymmetry algebra, the $V^{++}_5$ superfield is associated with the 
central charge generator $\hat{Z}$. Though the N=2 conformal SG itself has the 
vanishing central charge, N=2 matter hypermultiplets may have non-vanishing
central charges. It is natural to incorporate the central charge generator 
into the HSS covariant derivatives by redefining $\cd^{++}$ to $\cd_Z^{++}=
\cd^{++}+V^{++}_5\hat{Z}$, etc. At the component level the N=2 supermultiplet
 $V^{++}_5$ (in a WZ gauge) adds extra $8_B+8_F$ off-shell (field) degrees of 
freedom to the N=2 Weyl multiplet, thus forming together the so-called 
{\it minimal} off-shell N=2 SG multiplet with $32_B+32_F$ field components 
\cite{van}.

The extended HSS covariant derivative $\cd^{++}_Z$ also has to preserve 
analyticity, which implies a linear constraint on $V^{++}_5$. A solution to
the constraint reads \cite{kuz}
$$V^{++}_5=\fracm{i}{4}(D^+)^2\cg -\fracm{i}{4}(\bar{D}^+)^2\cg + 
v^{++}_5~,\eqno(A.10)$$
where $v^{++}_5$ is real analytic. The pre-gauge transformations (A.5) are to 
be appended by \cite{kuz}
$$ \d v^{++}_5=i(D^+)^4(\O^{--} -\sbar{\O}{}^{--})~. \eqno(A.11)$$
The related restrictions on the HSS parameter $\L_5$ in eq.~(A.9) are 
\cite{kuz} 
$$ \L_5=\fracm{i}{4}(D^+)^2l^{--}-\fracm{i}{4}
(\bar{D}^+)^2l^{--} +\l_5~,\eqno(A.12)$$
where $\l_5$ is real analytic. One easily finds 
$$ \d v^{++}_5 =-\l v^{++}_5  + \cd^{++}\l_5~.\eqno(A.13)$$ 

To compensate the remaining unwanted gauge symmetries (e.g., 
$SU(2)_{\rm conf.}$), one needs a {\it second compensator} that may have 
either a finite or the infinite number of the auxiliary field components. The 
three {\it minimal} formulations of N=2 Poincar\'e SG, each having 
$40_B+40_F$ off-shell (field) degrees of freedom, were described in 
ref.~\cite{van}. Unfortunately, all of them impose some restrictions on 
allowed N=2 matter couplings, which makes them of limited use in the context 
of generic N=2 NLSM. The most universal choice is given by 
a {\it real analytic} compensator $\o$ that compensates the N=2 local 
supersymmetry transformations of the analytic measure,
$$ \o'(\z',u') =Ber^{-1}\left(\fracmm{\pa(\z',u')}{\pa(\z,u)}\right)
\o(\z,u)~.\eqno(A.14)$$
It allows one to accommodate any N=2 NLSM in N=2 SG via the `covariantization' 
of the (rigid) N=2 NLSM action in HSS, by using the invariant analytic 
measure $d\z^{(-4)}du\,\o$. It is the absence of an analytic density that is 
responsible for the restricted N=2 matter couplings in the more conventional 
formulations of N=2 SG \cite{van}. Coupling to N=2 SG deformes hyper-K\"ahler 
geometry of the N=2 NLSM target space into quaternionic geometry 
\cite{bagw,hnlsm}. The actions of quaternionic NLSM coupled to N=2 
SG in HSS were recently investigated in ref.~\cite{iva}, where the density 
$\o$ was constructed in terms of a FS hypermultiplet superfield $q^+$ 
as $\o\sim (u^-_aq^{a+})^2$. When going in the opposite direction, a rigidly 
N=2 superconformal (special hyper-K\"ahler) NLSM arises from the quaternionic 
one after putting all the N=2 conformal SG fields to zero, $\cg=0$, together 
with the vanishing Maxwell multiplet, $v^{++}_5=0$, and $\o=1$.
\vglue.2in

\end{document}